\documentclass[prd,showpacs,showkeys,nofootinbib,floatfix,eqsecnum,fleqn,
                preprint,12pt,tightenlines]{revtex4} 

\usepackage{amsmath,amssymb,revsymb,graphicx,dcolumn}
\newcommand  {\version}{v6.0}

\newcommand{\beq}{\begin{equation}}
\newcommand{\eeq}{\end{equation}}
\newcommand{\beqa}{\begin{eqnarray}}
\newcommand{\eeqa}{\end{eqnarray}}
\newcommand{\bsubeqs}{\begin{subequations}}
\newcommand{\esubeqs}{\end{subequations}}

\newcommand{\half}{{\textstyle \frac{1}{2}}}    

\begin{document}

\noindent Phys. Rev. D 78, 063528 (2008)
\hfill    arXiv:0806.2805 [gr-qc] (\version)
\newline\vspace*{2mm}
\title{Dynamic vacuum variable and equilibrium approach in cosmology\vspace*{5mm}}
\author{F.R. Klinkhamer}
\email{frans.klinkhamer@physik.uni-karlsruhe.de}
\affiliation{\mbox{Institute for Theoretical Physics, University of Karlsruhe (TH),}\\
76128 Karlsruhe, Germany}
\author{G.E. Volovik}
\email{volovik@boojum.hut.fi}
\affiliation{\mbox{Low Temperature Laboratory, Helsinki University of Technology,}\\
P.O. Box 5100, FIN-02015 HUT, Finland\\
and\\
\mbox{L.D. Landau Institute for Theoretical Physics, Russian Academy of Sciences,}\\
Kosygina 2, 119334 Moscow, Russia}

\begin{abstract}
\vspace*{.5mm}\noindent
A modified-gravity theory is considered with a four-form
field strength $F$, a variable gravitational coupling parameter $G(F)$,
and a standard matter action. This theory provides a concrete
realization of the general vacuum variable $q$ as the four-form amplitude
$F$ and allows for a study of its dynamics.
The theory gives a flat Friedmann--Robertson--Walker
universe with rapid oscillations of the effective vacuum energy density
(cosmological ``constant''), whose amplitude drops to zero asymptotically.
Extrapolating to the present age of the Universe,
the order of magnitude of the average vacuum energy density agrees with
the observed near-critical vacuum energy density of the present universe.
It may even be that this type of oscillating vacuum energy density
constitutes a significant part of the so-called cold dark matter
in the standard Friedmann--Robertson--Walker framework.
\end{abstract}

\pacs{04.20.Cv, 98.80.Jk, 95.35.+d, 95.36.+x}
\keywords{general relativity, cosmology, dark matter, dark energy}
\maketitle

\section{Introduction}\label{sec:introduction}

In a previous article~\cite{KlinkhamerVolovik2008}, we proposed
to characterize a Lorentz-invariant quantum vacuum  by a nonzero conserved
relativistic ``charge'' $q$. This approach  allowed us to discuss the
\emph{thermodynamics} of the quantum vacuum, in particular,
thermodynamic properties as stability and compressibility.
We found that the vacuum energy density appears in two guises.

The \emph{microscopic} vacuum energy density is
characterized by an ultraviolet energy scale,
$\epsilon(q)\sim E^4_\text{UV}$.
For definiteness, we will take this energy scale $E_\text{UV}$ to be close to
the Planck energy scale $E_\text{Planck}\equiv \sqrt{\hbar\, c^5/G_\text{N}}
\approx 1.22 \times 10^{19}\,\text{GeV}$.
The \emph{macroscopic} vacuum energy density is, however,
determined by a particular thermodynamic quantity,
\mbox{$\widetilde{\epsilon}_\text{vac}(q) \equiv \epsilon - q\,d\epsilon/dq$,}
and it is this energy density that contributes to
the effective gravitational field equations at low energies.
For a self-sustained vacuum in full thermodynamic equilibrium and
in the absence of matter, the effective (coarse-grained) vacuum energy
density $\widetilde{\epsilon}_\text{vac}(q)$ is automatically nullified
(without fine tuning) by the spontaneous
adjustment of the vacuum variable $q$ to its equilibrium value $q_0$,
so that $\widetilde{\epsilon}_\text{vac}(q_0)=0$.
This implies that the effective cosmological
constant $\Lambda$ of a perfect quantum vacuum is strictly zero,
which is consistent with the requirement of Lorentz invariance.

The presence of thermal matter makes the vacuum state Lorentz noninvariant
and leads to a readjustment of the variable $q$ to a new equilibrium value,
$q_0^\prime = q_0 + \delta q$,
which shifts the effective vacuum energy density away from zero,
$\widetilde{\epsilon}_\text{vac}(q_0 + \delta q) \ne 0$.
The same happens with other
types of perturbations that violate Lorentz invariance,
such as the existence of a spacetime boundary or an interface.
According to this approach, the present value of
$\widetilde{\epsilon}_\text{vac}$ is nonzero but small because
the universe is close to equilibrium and Lorentz-noninvariant
perturbations of the quantum vacuum are small (compared with the
ultraviolet scale which sets the microscopic energy density $\epsilon$).

The situation is different for Lorentz-invariant perturbations of the vacuum,
such as the formation of scalar condensates as discussed in
Ref.~\cite{KlinkhamerVolovik2008} or quark/gluon condensates
derived from quantum chromodynamics (cf. Ref.~\cite{Shifman-etal1978}).
In this case,
the variable $q$ shifts in such a way that it completely compensates
the energy density of the perturbation
and the effective cosmological constant is again zero
in the new Lorentz-invariant equilibrium vacuum.

The possible origin of the conserved vacuum charge $q$
in the perfect Lorentz-invariant quantum vacuum
was discussed in Ref.~\cite{KlinkhamerVolovik2008}  in general terms.
But a specific example was also given in terms of  a four-form field
strength $F$~\cite{DuffNieuwenhuizen1980,Aurilia-etal1980,Hawking1984+Duff+Wu,
DuncanJensen1989,BoussoPolchinski2000,Aurilia-etal2004}.
Here, we use this explicit realization with a
four-form field $F$ to study the \emph{dynamics} of
the vacuum energy, which describes the relaxation of the
vacuum energy density $\widetilde{\epsilon}_\text{vac}$
(effective cosmological ``constant'') from its  natural Planck-scale value
at early times to a naturally small value at late times.
In short, the present cosmological constant is small because
the Universe happens to be old.\footnote{An extensive but nonexhaustive
list of references to research papers
and reviews on the so-called ``cosmological constant problem(s)''
can be found in Ref.~\cite{KlinkhamerVolovik2008}.
A recent review on cosmic ``dark energy''  is given in
Ref.~\cite{Frieman-etal2008}.}

The results of the present article show that, for the type of theory considered,
the decay of $\widetilde{\epsilon}_\text{vac}$ is accompanied by rapid
oscillations of the vacuum variable $F$ and that the relaxation of
$\widetilde{\epsilon}_\text{vac}$ mimics the behavior of cold dark matter
(CDM) in a standard Friedmann--Robertson--Walker (FRW) universe. This
suggests that part of the inferred CDM may come from dynamic
vacuum energy density and may also give a clue
to the solution of the so-called coincidence problem
\cite{Frieman-etal2008}, namely, why the approximately constant vacuum
energy density is precisely now of the same order as the time-dependent
CDM energy density.

These results are obtained by the following steps.
In Sec.~\ref{sec:Gravity-with-F-field}, a modified-gravity theory with
a four-form field $F$ is defined in terms of general functions for the
microscopic energy density $\epsilon(F)$ and variable gravitational
coupling parameter $G(F)$.
In Sec.~\ref{sec:de-Sitter-expansion}, the dynamics of the corresponding
de-Sitter universe without matter is discussed and,
in Sec.~\ref{sec:FRW universe}, the dynamics of a flat FRW
universe with matter, using simple \emph{Ans\"{a}tze} for the functions
$\epsilon(F)$ and $G(F)$.
In Sec.~\ref{sec:Equilibrium-approach}, the approach to equilibrium in such
a FRW universe is studied in detail and the above mentioned vacuum
oscillations are established.
In Sec.~\ref{sec:Conclusion}, the main results are summarized.

\section{Gravity with $\boldsymbol{F}$ field and
         variable gravitational coupling}
\label{sec:Gravity-with-F-field}

Here, and in the following, the vacuum variable $q$ is represented by a
four-form field $F$.\footnote{To clarify our notation,
a four-form field has components
$F_{\kappa\lambda\mu\nu}(x)$ which can always be written
as $e_{\kappa\lambda\mu\nu}\,\sqrt{|g(x)|} \,F(x)$,
in terms of the constant Levi--Civita symbol $e_{\kappa\lambda\mu\nu}$,
the determinant of the metric $g(x)\equiv \det g_{\mu\nu}(x)$,
and a real scalar field $F(x)$. Hence, we can simply write
$F$ if we speak about the four-form field.
However, this scalar field $F(x)$ is not fundamental,
as will become clear later.}
The corresponding action is given by a generalization
of the action in which only a quadratic function of $F$ is used (see, e.g.,
Refs.~\cite{DuffNieuwenhuizen1980,Aurilia-etal1980,Hawking1984+Duff+Wu,
DuncanJensen1989,BoussoPolchinski2000,Aurilia-etal2004}).
Such a quadratic function gives rise to a gas-like
vacuum~\cite{KlinkhamerVolovik2008}.
But a gas-like vacuum cannot exist in equilibrium without external
pressure, as the equilibrium vacuum charge vanishes, $q_0=0$.
A self-sustained vacuum requires a more complicated
function $\epsilon(F)$ in the action, so that the equilibrium at
zero external pressure occurs for $q_0\neq 0$. An example of an appropriate
function $\epsilon(F)$ will be given in Sec.~\ref{sec:Vacuum-energy}.

The action is chosen as in Ref.~\cite{KlinkhamerVolovik2008}
but with one important modification:
Newton's constant $G_\text{N}$ is replaced by
a gravitational coupling parameter $G$ which is taken to
depend on the state of the vacuum and thus on the vacuum variable $F$.
Such a $G(F)$ dependence is natural and must, in principle,
occur in the quantum vacuum. Moreover, a $G(F)$ dependence allows
the cosmological ``constant'' to change with time, which is otherwise
prohibited by the Bianchi identities
and energy-momentum conservation~\cite{Weinberg1972,Wald1984}.

Specifically, the action considered takes the following form ($\hbar=c=1$):
\bsubeqs\label{eq:EinsteinF-all}
\beqa
S[A, g,\psi]&=& -
\int_{\mathbb{R}^4} \,d^4x\, \sqrt{|g|}\,\left(\frac{R}{16\pi G(F)} +
\epsilon(F)+\mathcal{L}^\text{M}(\psi)\right)\,,
\label{eq:actionF}\\[2mm]
F^2 &\equiv& - \frac{1}{24}\, F_{\kappa\lambda\mu\nu}\,F^{\kappa\lambda\mu\nu}\,,\quad
F_{\kappa\lambda\mu\nu}\equiv \nabla_{[\kappa}A_{\lambda\mu\nu]}\,,
\label{eq:Fdefinition}\\[2mm]
F_{\kappa\lambda\mu\nu}&=&F\,e_{\kappa\lambda\mu\nu}\,\sqrt{|g|}\,,
\quad
F^{\kappa\lambda\mu\nu}=F \,e^{\kappa\lambda\mu\nu}/\sqrt{|g|}\,,
\label{eq:Fdefinition2}
\eeqa
\esubeqs
where $\nabla_\mu$ denotes a covariant derivative
and a square bracket around spacetime indices complete antisymmetrization.
The functional dependence on $g$ has been kept
implicit on the right-hand side of \eqref{eq:actionF}, showing only the
dependence on $F=F(A,g)$ and $\psi$.
The field $\psi$ in \eqref{eq:actionF} stands, in fact,
for a generic low-energy matter field
with a scalar Lagrange density, $\mathcal{L}^\text{M}(\psi)$,
which is assumed to be without $F$--field dependence
(this assumption can be relaxed later by changing the low-energy
constants in $\mathcal{L}^\text{M}$ to $F$--dependent parameters).
It is also assumed that a possible constant term $\Lambda^\text{M}$ in
$\mathcal{L}^\text{M}(\psi)$ has been absorbed in $\epsilon(F)$,
so that, in the end, $\mathcal{L}^\text{M}(\psi)$ contains only
$\psi$--dependent terms.
In this section, the low-energy fields are indicated by lower-case
letters, namely, $g_{\mu\nu}(x)$ and $\psi(x)$,
whereas the fields originating from the microscopic theory
are indicated by upper-case letters, namely,
$A(x)$ and $F(x)$ [later also $\Phi(x)$].
Throughout, we use the conventions of Ref.~\cite{Weinberg1972}, in particular,
those for the Riemann tensor and the metric signature $(-+++)$.

The variation of the action \eqref{eq:actionF} over the three-form
gauge field $A$ gives the generalized Maxwell equation,
\begin{equation}
\nabla_\nu \left(\sqrt{|g|} \;\frac{F^{\kappa\lambda\mu\nu}}{F} \left(
\frac{d\epsilon(F)}{d F}+\frac{R}{16\pi} \frac{dG^{-1}(F)}{d F} \right)
\right)=0\,,
\label{eq:Maxwell}
\end{equation}
and the variation over the metric $g_{\mu\nu}$ gives
the generalized Einstein equation,
\begin{eqnarray}
&&
\frac{1}{8\pi G(F)}
\left( R_{\mu\nu}-\frac{1}{2}\,R\,g_{\mu\nu}\right)
+\frac{1}{16\pi}\, F\,\frac{d G^{-1}(F)}{d F}\, {R}\,g_{\mu\nu}
\nonumber\\[2mm]
&&+ \frac{1}{8\pi} \Big( \nabla_\mu\nabla_\nu\, G^{-1}(F) - g_{\mu\nu}\,
\Box\, G^{-1}(F)\Big) -\widetilde{\epsilon}(F) g_{\mu\nu}
+T^\text{M}_{\mu\nu} =0\,, \label{eq:EinsteinEquationF}
\end{eqnarray}
where $\Box$ is the invariant d'Alembertian,
$T^\text{M}_{\mu\nu}$ the energy-momentum tensor of the matter
field $\psi$, and $\widetilde{\epsilon}$ the effective
vacuum energy density
\begin{equation}
\widetilde{\epsilon}(F)\equiv \epsilon(F) -F\,\frac{d\epsilon(F)}{d F}\,,
\label{eq:widetilde-epsilon}
\end{equation}
whose precise form has been argued on thermodynamic grounds
in Ref.~\cite{KlinkhamerVolovik2008}.

At this point  two remarks may be helpful.
First, observe that the action \eqref{eq:actionF} is not quite the one of
Brans--Dicke theory~\cite{Weinberg1972,BransDicke1961}, as
the argument of $G(F)$ is not a fundamental scalar field
but involves the inverse metric [needed to change the covariant tensor
$F_{\kappa\lambda\mu\nu}$ into a contravariant tensor $F^{\kappa\lambda\mu\nu}$
for the definition of $F \equiv \sqrt{F^2}$ according to
\eqref{eq:Fdefinition}]. This implicit metric dependence of $G(F)$
explains the origin of the second term
on the left-hand side of \eqref{eq:EinsteinEquationF}.
Second, observe that the three-form gauge field $A$ does not
propagate physical degrees of freedom in flat spacetime
\cite{DuffNieuwenhuizen1980,Aurilia-etal2004}.
Still, $A$ has gravitational effects, both classically
in the modified-gravity theory with $G=G(F)$
as discussed in the present article
(see, in particular, Sec.~\ref{sec:Effective-CDM-like-behavior})
and quantum-mechanically already in the standard gravity theory
with $G=G_\text{N}$ (giving, for example, a nonvanishing
gravitational trace anomaly~\cite{DuffNieuwenhuizen1980}).

Using \eqref{eq:Fdefinition2} for $F^{\kappa\lambda\mu\nu}$,
we obtain the Maxwell equation \eqref{eq:Maxwell} in the form
\begin{equation}
\partial_\nu \left( \frac{d\epsilon(F)}{d F}+\frac{R}{16\pi}
\frac{dG^{-1}(F)}{d F} \right) =0\,.
\label{eq:Maxwell2}
\end{equation}
The solution is simply
\begin{equation}
 \frac{d\epsilon(F)}{d F}+\frac{R}{16\pi} \frac{dG^{-1}(F)}{d F} =\mu \,,
\label{eq:MaxwellSolution}
\end{equation}
with an integration constant $\mu$. Hence, the constant $\mu$
is seen to emerge dynamically.
In a thermodynamic equilibrium state, this constant becomes a genuine
chemical potential corresponding to the conservation law obeyed by
the vacuum ``charge'' $q\equiv F$. Indeed, the integration constant $\mu$ is,
according to \eqref{eq:MaxwellSolution},  thermodynamically conjugate to $F$
in an equilibrium state with vanishing Ricci scalar $R$.

Eliminating $dG^{-1}/dF$ from \eqref{eq:EinsteinEquationF} by use of
\eqref{eq:MaxwellSolution}, the generalized Einstein equation becomes
\begin{equation}
\frac{1}{8\pi G(F)}\Big( R_{\mu\nu}-\half\,R\,g_{\mu\nu} \Big) + \frac{1}{8\pi}
\Big( \nabla_\mu\nabla_\nu\, G^{-1}(F) - g_{\mu\nu}\, \Box\, G^{-1}(F)\Big)
-\Big(\epsilon(F)-\mu\, F \Big)\, g_{\mu\nu}+T^\text{M}_{\mu\nu} =0\,,
\label{eq:EinsteinEquationF2}
\end{equation}
which will be used in the rest of this article,
together with \eqref{eq:MaxwellSolution}.

Equations \eqref{eq:MaxwellSolution} and \eqref{eq:EinsteinEquationF2}
can also be obtained if we use, instead of the original action,
an effective action in terms of a Brans--Dicke-type scalar field
$\Phi(x)$ with mass dimension 2, setting $\Phi(x)\to F(x)$ afterwards.
Specifically, this effective action is given by
\beq
 S_\text{eff}[\Phi,\mu, g,\psi] = -
\int_{\mathbb{R}^4} \,d^4x\, \sqrt{|g|}\,\left(\frac{R}{16\pi G(\Phi)} +
\big(\epsilon(\Phi)-\mu \,\Phi\big) +\mathcal{L}^\text{M}(\psi)\right)\,.
\label{eq:actionF-muform}
\eeq
The potential term in \eqref{eq:actionF-muform} contains,
different from a conventional Brans--Dicke potential $V(\Phi)$,
a linear term, $-\mu \,\Phi$, for a constant $\mu$ of mass dimension 2.
This linear term reflects the
fact that our effective scalar field $\Phi$ is not an arbitrary field
but should be a conserved quantity, for which the constant parameter
$\mu$ plays the role of a chemical potential that is thermodynamically
conjugate to $\Phi$.

Indeed, if $\Phi$ in \eqref{eq:actionF-muform} is replaced by a four-form field $F$
given in terms of the three-form potential $A$,
the resulting $\mu F$ term in the effective action does not contribute
to the equations of motion \eqref{eq:Maxwell},
because it is a total derivative,
\begin{equation}
\int_{\mathbb{R}^4} \,d^4x\; \sqrt{|g|}\, \mu\, F = -
\frac{\mu}{24} ~e^{\kappa\lambda\mu\nu}
\int_{\mathbb{R}^4} \,d^4x\; F_{\kappa\lambda\mu\nu} \,,
\label{eq:actionSurface}
\end{equation}
where the constant $\mu$ plays the role of a Lagrange multiplier
related to the conservation of vacuum ``charge'' $F$
(see also the discussion in Refs.~\cite{Aurilia-etal1980,DuncanJensen1989},
where $\mu$ is compared with the $\theta$ parameter of quantum chromodynamics).

Instead of the large microscopic energy density $\epsilon(F)$
in the original action \eqref{eq:actionF},
the potentially smaller macroscopic vacuum energy density
$\rho_\text{V}\equiv \epsilon(F)-\mu F$ enters the effective
action \eqref{eq:actionF-muform}.
Precisely this macroscopic vacuum energy density gravitates and
determines the cosmological term in the gravitational field equations
\eqref{eq:EinsteinEquationF2}.

Equations \eqref{eq:MaxwellSolution} and \eqref{eq:EinsteinEquationF2}
are universal: they do not depend on the particular origin of the vacuum field $F$.
The $F$ field can be replaced by any conserved variable $q$,
as discussed in Ref.~\cite{KlinkhamerVolovik2008}.
Observe that, for thermodynamics, the parameter $\mu$ is the quantity
that is thermodynamically conjugate to $q$  and that, for dynamics,
$\mu$  plays the role of a  Lagrange multiplier.
The functions $\epsilon(q)$ and $G(q)$ can be considered
to be phenomenological parameters in an effective low-energy theory
(see also the general discussion in the Appendix of Ref.~\cite{Volovik2006}).

Before we turn to the cosmological solutions of our particular
$F$ theory \eqref{eq:EinsteinF-all}, it may be useful
to mention the connection with so-called $f(R)$ models
which have recently received considerable attention
(see, e.g., Refs.~\cite{Starobinsky2007,SotiriouFaraoni2008}
and references therein).
The latter are purely phenomenological models,
in which the linear function of the Ricci scalar $R$
from the Einstein--Hilbert action term
is replaced by a more general function $f(R)$.
This function $f(R)$ can, in principle, be
adjusted to fit the astronomical observations
and to produce a viable cosmological model.
Returning to our $F$ theory, we can
express $F$ in terms of $R$ by use of \eqref{eq:MaxwellSolution} and
substitute the resulting expression $F(R)$ into \eqref{eq:EinsteinEquationF2}.
This gives an equation for the metric field, which
is identical to the one of $f(R)$ cosmology.
(The latter result is not altogether surprising as the metric $F(R)$ model
is known to be equivalent to a Brans--Dicke model without kinetic term
\cite{SotiriouFaraoni2008} and the same holds for our effective
action \eqref{eq:actionF-muform} at the classical level.)
In this way, the $F$ theory introduced in this section
(or, more generally, $q$ theory as mentioned in the previous paragraph)
may give a microscopic justification for the phenomenological $f(R)$ models
used in theoretical cosmology and may allow for a choice between different
classes of model functions $f(R)$ based on fundamental physics.

\section{de-Sitter expansion}
\label{sec:de-Sitter-expansion}

Let us, first, consider \emph{stationary} solutions of the generalized
Maxwell--Einstein equations from the effective action \eqref{eq:actionF-muform}.
At this moment, we are primarily interested in the class of
spatially flat, homogeneous, and isotropic universes.
In this class,  only the matter-free de-Sitter universe is stationary.

The de-Sitter universe is characterized by a time-independent
Hubble parameter $H$ (that is, a genuine Hubble constant $H$),
which allows us to regard this universe as a thermodynamic equilibrium system.
Using
\begin{equation}
R_{\mu\nu}=\frac{1}{4}\,g_{\mu\nu}\, R\,,\quad
R=-12\: H^2 \,,
\label{eq:deSitter}
\end{equation}
we get from \eqref{eq:MaxwellSolution} and \eqref{eq:EinsteinEquationF2}
two equations for the constants $F$ and $H$:
\bsubeqs\label{eq:EquationFmu}
\beqa
\Big(\frac{d\epsilon(F)}{d F}-\mu \Big)
&=&
\frac{3 H^2}{4\pi }\, \frac{dG^{-1}(F)}{d F}\,,
\label{eq:EquationFmu-rhoV}
\\[2mm]
\Big( \epsilon(F)-\mu\, F\Big)
&=&
\frac{3 H^2}{8\pi}\,G^{-1}(F)\,,
\label{eq:EquationFmu-drhoVdF}
\eeqa
\esubeqs
with $\mu$ considered given.

Eliminating the chemical potential $\mu$ from the above equations,
we find the following equation for $F$:
\beq
\widetilde{\epsilon}(F)\equiv \epsilon(F)-F\,\frac{d\epsilon(F)}{d F}=
\frac{3H^2}{8\pi} \left(G^{-1}(F)-2F\,\frac{dG^{-1}(F)}{d F}\right) \,,
\label{eq:EquationF}
\eeq
where the functions $\epsilon(F)$ and $G^{-1}(F)$ are assumed to be known.

The perfect quantum vacuum corresponds to $H=0$ and describes Minkowski
spacetime. The corresponding equilibrium values $F=F_0$  and $\mu=\mu_0$
in the perfect quantum vacuum are determined from the following equations:
\begin{equation}
\epsilon(F_0)-F\,\frac{d\epsilon(F)}{dF}\,\Bigg|_{F=F_0}
=0\,,\quad\mu_0=\frac{d\epsilon(F)}{dF}\,\Bigg|_{F=F_0}\,,
\label{eq:EquilibriumF0}
\end{equation}
which are obtained from
\eqref{eq:MaxwellSolution} and \eqref{eq:EinsteinEquationF2} by recalling
that the perfect quantum vacuum is the equilibrium vacuum in the
absence of matter and gravity fields ($T^\text{M}_{\mu\nu} =R=0$).

If $H$ is nonzero but small compared with the Planck energy scale,
the $H^2$ term on the right-hand side of \eqref{eq:EquationF}
can be considered as  a perturbation.
Then, the correction $\delta F=F-F_0$ due to the expansion is given by
\begin{equation}
\frac{\delta F}{F_0} =-\frac{3}{8\pi} \,\chi(F_0)\; H^2\left(G^{-1}(F_0) -
2 F\,\frac{dG^{-1}(F)}{d F}\,\Bigg|_{F=F_0}\,\right)\,,
\label{eq:CorrectionToF}
\end{equation}
where $\chi(F_0)$ is the vacuum compressibility introduced in
Ref.~\cite{KlinkhamerVolovik2008},
\begin{equation}
\chi(F_0)\equiv
\left(F^2\:\frac{d^2\epsilon(F)}{dF^2}\,\Bigg|_{F=F_0}\right)^{-1}\,.
\label{eq:chi}
\end{equation}
Equally, the chemical potential is modified by the expansion ($H\ne 0$):
\begin{equation}
\mu=\mu_0+\delta\mu=\frac{d\epsilon(F)}{d F}\,\Bigg|_{F=F_0} -
\frac{3 H^2}{8\pi G(F_0)F_0} \,.
\label{eq:ChemicalPotential}
\end{equation}

But, instead of fixing $H$, it is also possible to fix the integration constant
$\mu$.  From \eqref{eq:EquationFmu}, we then obtain the other parameters as
functions of $\mu$: $H(\mu)$, $F(\mu)$, and
$\rho_\text{V}(\mu) \equiv \epsilon(\mu)-\mu F(\mu)$.
The cosmological constant $\Lambda(\mu)\equiv \rho_\text{V}(\mu)$
is zero for $\mu=\mu_0$,
which corresponds to thermodynamic equilibrium in the absence of
external pressure and expansion
[$P_\text{external}=P_\text{vac}(\mu_0)=-\Lambda(\mu_0)=0$]. From now on,
the physical situation considered will be the one determined
by having a fixed chemical potential $\mu$.

The de-Sitter universe is of interest because it is an equilibrium system
and, therefore, may serve as the final state of a dynamic universe
with matter included (see Sec.~\ref{sec:Equilibrium-approach}).

\section{Dynamics of a flat FRW universe}
\label{sec:FRW universe}

\subsection{General equations}
\label{sec:Dynamical-equations}

The discussion of this section and the next is restricted
to a spatially flat FRW universe, because of two reasons.
The first reason is that
flatness is indicated by the data from observational cosmology
(cf. Refs.~\cite{Frieman-etal2008,Mukhanov2005,Weinberg2008,
                 Astier-etal2006,Riess-etal2007,Komatsu-etal2008}
and references therein).
The second reason is that flatness is a natural property
of the quantum vacuum in an emergent gravity theory
(cf. Ref.~\cite{KlinkhamerVolovik2008} and references therein).
In addition, the matter energy-momentum tensor for the model universe
is taken as that of a perfect fluid characterized by the energy density
$\rho_\text{M}$ and isotropic pressure $P_\text{M}$.
As mentioned in the previous section, the physics of the $F$ field
is considered to be specified by a fixed chemical potential $\mu$.

For a spatially flat ($k=0$) FRW universe~\cite{Weinberg1972}
with expansion factor $a(t)$, the  homogenous matter has, in general, a
time-dependent energy density $\rho_\text{M}(t)$ and pressure $P_\text{M}(t)$.
Equally, the scalar field entering the four-form field-strength
tensor \eqref{eq:Fdefinition2} is taken to be homogenous and time dependent,
$F_{\kappa\lambda\mu\nu}=F(t)\,|a(t)|^3 \,e_{\kappa\lambda\mu\nu}\,$.

With a time-dependent Hubble parameter $H(t)\equiv (da/dt)/a$,
we then have from the reduced Maxwell equation \eqref{eq:MaxwellSolution}:
\begin{equation}
\frac{3}{8\pi} \frac{dG^{-1}}{d F} \left(\frac{d H}{dt} +2H^2 \right) =
 \frac{d\epsilon}{d F}-\mu \,,
\label{eq:MaxwellFRW}
\end{equation}
and from the Einstein equation \eqref{eq:EinsteinEquationF2}:
\bsubeqs\label{eq:EinsteinFRW-all}
\beqa H^2 &=&
\frac{8\pi}{3}\, G\, \rho_\text{tot} - H G \,\frac{d G^{-1}}{dt}\,,
\label{eq:EinsteinFRW-E00eq}\\[2mm]
2\,\frac{d H}{dt}+3H^2 &=& -8\pi\, G\, P_\text{tot} -  2HG \,\frac{d G^{-1}}{dt}
      -G \,\frac{d^2 G^{-1}}{dt^2}  \,,
\label{eq:EinsteinFRW-E11eq}
\eeqa
\esubeqs
with total energy density and
pressure \beq\label{eq:rho-total} \rho_\text{tot}\equiv
\rho_\text{V}+\rho_\text{M}\,,\quad P_\text{tot} \equiv
P_\text{V}+P_\text{M}\,, \eeq for the effective vacuum energy density \beq
\rho_\text{V}(F)= -P_\text{V}(F) = \epsilon(F) -\mu F\,.
\label{eq:EinsteinFRW-rhoV} \eeq

With definition \eqref{eq:EinsteinFRW-rhoV}, the reduced
Maxwell equation \eqref{eq:MaxwellFRW} can be written as
\begin{equation}
\dot\rho_\text{V} =
\frac{3}{8\pi}\, \frac{dG^{-1}}{dt} \left(\dot{H} +2H^2\right)  \,,
\label{eq:MaxwellFRW-a-deriv}
\end{equation}
where the overdot stands for differentiation with respect to cosmic time $t$.
The above equations give automatically energy-conservation of matter,
\beq\label{eq:matter energy-conservation}
\dot{\rho}_\text{M} +3H\, \Big(P_\text{M}+\rho_\text{M}\Big) = 0 \,,
\eeq
as should be the case for a standard matter field $\psi$
(recall that $\nabla^\mu\,T^\text{M}_{\mu\nu}=0$ follows from the
invariance of $S^\text{M}[g_{\mu\nu},\psi]$ under general
coordinate transformations; cf. Appendix E of Ref.~\cite{Wald1984}).

\subsection{Model for $\boldsymbol{\epsilon(F)}$}
\label{sec:Vacuum-energy}

The equations of Sec.~\ref{sec:Dynamical-equations}
allow us to study the development of the Universe
from very small (near-Planckian) time scales to macroscopic time scales.
Because the results do not depend very much on the details
of the functions  $\epsilon(F)$ and  $G(F)$, it is possible
to choose the simplest functions for an exploratory investigation.
The only requirements are that the vacuum is self-sustained
[i.e., \eqref{eq:EquilibriumF0} has a solution with nonzero $F_0$] and
that the vacuum is stable
[i.e., the vacuum compressibility \eqref{eq:chi} is positive,
$\chi(F_0)>0$].

A simple choice for the function $\epsilon(F)$ is
\begin{equation}
\epsilon(F)=\frac{1}{2\chi}\left(-\frac{F^2}{F_0^2}
+\frac{F^4}{3F_0^4}\right)\,,
\label{eq:epsilonF-Ansatz}
\end{equation}
where $\chi>0$ is a constant parameter (vacuum compressibility) and $F_0$
the value of $F$ in a particular equilibrium vacuum satisfying
\eqref{eq:EquilibriumF0}. The equilibrium value of the chemical potential
$\mu$ in the perfect vacuum is then given by
\begin{equation}
\mu_0=-\frac{1}{3\chi F_0}\,.
\label{eq:mu-Ansatz}
\end{equation}

The microscopic parameters $F_0$ and $\chi$ are presumably
determined by the Planck energy scale, $|F_0|\sim E^2_\text{Planck}$
and $\chi\sim 1/\epsilon(F_0) \sim 1/ E^4_\text{Planck}$.
 From \eqref{eq:mu-Ansatz}, we then see that $|\mu_0|\sim |F_0|$.
Let us now rewrite our equations in microscopic (Planckian) units by
introducing appropriate dimensionless variables
$f$, $y$, $u$, $k$, $h$, and $\tau$:
\bsubeqs\label{eq:Dimensionless1}
\beqa
F&=&f F_0\,,\quad y \equiv f-1\,,
\label{eq:Dimensionless1-f-y}
\\[2mm]
\mu&=&\frac{u}{\chi F_0}\,,\quad G^{-1}(F)=k(f) |F_0|\,,
\label{eq:Dimensionless1-u-k}
\\[2mm]
H&=&h/\sqrt{\chi  |F_0|}\,,\quad t=\tau \,\sqrt{\chi  |F_0|}\,,
\label{eq:Dimensionless1-h-tau}
\eeqa
\esubeqs
where the variable $y$ has been introduced in anticipation of the
calculations of Sec.~\ref{sec:Equilibrium-approach}.
The corresponding normalized vacuum and matter energy densities are
defined as follows:
\beq
\rho_\text{V,M} =\frac{r_\text{V,M}}{ \chi}  \,,
\label{eq:DimensionlessVMenergies}
\eeq
and \emph{Ansatz} \eqref{eq:epsilonF-Ansatz}  gives
\beq
 r_\text{V}      = \frac{1}{2}\left( -f^2+\frac{1}{3}f^4\right)-u f\,,
\label{eq:Dimensionless-eV}
\eeq
with $u=u_0=-1/3$ from \eqref{eq:mu-Ansatz}.

 From the Maxwell equation \eqref{eq:MaxwellFRW},
 the Friedmann equation \eqref{eq:EinsteinFRW-E00eq},
and the matter conservation equation
\eqref{eq:matter energy-conservation}, we finally obtain a closed system
of three ordinary differential equations (ODEs)
for the three dimensionless variables $h$, $f$, and $r_\text{M}$:
\bsubeqs\label{eq:3eqsFRWdim} \beqa
\frac{3}{8\pi}\, \frac{d k}{d f} \left(\frac{d h}{d\tau} +2h^2 \right) &=&
 \frac{dr_\text{V}}{d f} \,,
\label{eq:MaxwellFRWdim}\\[2mm]
\frac{3}{8\pi}\,\left(
h\,\frac{d k}{d f}\,\frac{d f}{d\tau}+ k\,h^2 \right)&=& r_\text{V}+r_\text{M} \,,
\label{eq:EinsteinFRWdim}\\[2mm]
\frac{dr_\text{M}}{d\tau}+3h \,\big(1+w_\text{M}\big)\,\,r_\text{M}&=& 0\,,
\label{eq:ConservationFRWdim}
\eeqa
\esubeqs
with matter equation-of-state (EOS) parameter
$w_\text{M}\equiv P_\text{M}/\rho_\text{M}$.

\subsection{Model for $\boldsymbol{G(F)}$}
\label{sec:G(F)}

Next, we need an appropriate \emph{Ansatz} for the function $G(F)$
or the dimensionless function $g(f) \equiv 1/k(f)$ in microscopic units.
There are several possible types of behavior for $G(F)$, but we
may reason as follows.

It is possible that for $F^2\ll F_0^2$ (i.e., in the gas-like vacuum)
the role of the Planck scale is played by
$E_\text{P}(F)\equiv |\epsilon(F)|^{1/4}\sim |F|^{1/2}$.
The gravitational coupling parameter would then be given by
\begin{equation}
\frac{1}{G(F)}\sim E^2_\text{P}(F)\sim |F|~\,,\quad |F|\ll |F_0|\,.
\label{eq:NewtonF}
\end{equation}
This equation also gives the correct estimate for $G(F)$ in
the equilibrium vacuum: $1/G(F_0)\sim E_\text{Planck}^2(F_0) \sim |F_0|$,
according to the estimates given a few lines below \eqref{eq:mu-Ansatz}.
Thus, a simple choice for the function $G^{-1}(F)$ is
\begin{equation}
G^{-1}(F)=s\, |F|\,,\quad  k(f)=s\, f\,,
\label{eq:Ginverse-Ansatz}
\end{equation}
with $f$ taken positive (in fact, $f\sim 1$ for $F\sim F_0$) and a single
time-independent dimensionless parameter $s$ also taken positive.

Assuming \eqref{eq:Ginverse-Ansatz}, the three ODEs \eqref{eq:3eqsFRWdim} become
\bsubeqs\label{eq:3eqsFRWdim2}
\beqa
\sigma \left(\frac{d h}{d\tau} +2h^2 \right) &=& \frac{dr_\text{V}}{d f} \,,
\label{eq:MaxwellFRWdim2}
\\[2mm]
\sigma\left(h \,\frac{d f}{d\tau}+f h^2\right)&=&r_\text{V}+r_\text{M}\,,
\label{eq:EinsteinFRWdim2}
\\[2mm]
  \frac{dr_\text{M}}{d\tau}+3h\,\big(1+w_\text{M}\big)\,r_\text{M}&=& 0\,,
\label{eq:ConservationFRWdim2}
\eeqa
\esubeqs
with $r_\text{V}=r_\text{V}(f)$ given by \eqref{eq:Dimensionless-eV}
and a single free parameter $\sigma\equiv 3s/8\pi$.
This dimensionless parameter $\sigma$ is of order $1$
if  the physics of $F$ field is solely determined by the Planck energy scale
(i.e., for  $F_0^2\sim 1/\chi  \sim \mu_0^2\sim  E_\text{Planck}^4$).
Anyway, the parameter $\sigma$ can be absorbed in $h$ and $\tau$ by the
redefinition $h\rightarrow h/\sqrt{\sigma}$ and
$\tau \rightarrow \tau\sqrt{\sigma}$. Henceforth, we set $\sigma=1$
in \eqref{eq:3eqsFRWdim2}, so that there are no more free parameters
except for the EOS parameter $w_\text{M}$ (taken to be time independent
in the analysis of the next section).

\section{Equilibrium approach in a flat FRW universe}
\label{sec:Equilibrium-approach}

\subsection{Equations at the equilibrium point $\mu=\mu_0$}
\label{sec:Equations_mu0}

Equations \eqref{eq:MaxwellFRWdim2}--\eqref{eq:ConservationFRWdim2}
allow us to study the evolution of the flat FRW universe towards a
stationary state, if the initial universe was far away from equilibrium.
The final state can be either the de-Sitter universe of
Sec.~\ref{sec:de-Sitter-expansion} with $\rho_\text{M} =0$
and $\rho_\text{V} \ne 0$
or the perfect quantum vacuum (Minkowski spacetime)
with $H=\rho_\text{M} =\rho_\text{V} =0$
and $f=1$. Here, we consider the latter possibility where the system
approaches one of the two perfect quantum vacuum states
with $f=1$, which correspond to  either $F=+|F_0|$ or $F=-|F_0|$
for vacuum energy density \eqref{eq:epsilonF-Ansatz}.

Such an equilibrium vacuum state can be reached only if the
chemical potential $\mu$ corresponds to full equilibrium: $\mu=\mu_0$
as given by \eqref{eq:mu-Ansatz}
or $u=u_0=-1/3$ in microscopic units \eqref{eq:Dimensionless1-u-k}.
Since $\mu$ is an integration constant,
there may be a physical reason for the special value $\mu_0$.
Indeed, the starting nonequilibrium state could, in turn, be obtained
by a large perturbation of an initial equilibrium vacuum.
In this case, the integration constant would
remember the original perfect equilibrium.
(The evolution towards a de-Sitter universe for $\mu \neq \mu_0$ will be
only briefly discussed in Sec.~\ref{sec:Numerical results}.)

In order to avoid having to consider quantum corrections
to the Einstein equation, which typically appear near the time $\tau \sim 1$
(or $t\sim t_\text{Planck}\equiv\hbar/E_\text{Planck}$), we consider
times $\tau \gg 1$, where the quantum  corrections can be expected to be small.
For these relatively large times, $f$ is close to unity and we
may focus on the deviation from equilibrium as given by the
variable $y$ defined in \eqref{eq:Dimensionless1-f-y}.

Taking the time derivative of \eqref{eq:EinsteinFRWdim2} for $\sigma=1$
and using \eqref{eq:MaxwellFRWdim2} and \eqref{eq:ConservationFRWdim2},
we obtain
\begin{equation}
\ddot{y} -\dot{y} h +2(1+y) \dot{h}=-3\,\big(1+w_\text{M}\big)\,r_\text{M}\,,
\label{eq:EinsteinDerivative}
\end{equation}
where, from now on, the overdot stands for differentiation with respect
to $\tau$. Next, eliminate the matter density $r_\text{M}$ from equations
\eqref{eq:EinsteinFRWdim2} and  \eqref{eq:EinsteinDerivative}, in order to
obtain a system of two equations for the two variables $y$ and $h$:
\bsubeqs\label{eq:Matter excluded1and2}
\beqa
\ddot{y} -\dot{y} h +2(1+y) \dot{h} &=&
-3\,\big(1+w_\text{M}\big)\,\left[\dot{y} h + (1+y)h^2 - r_\text{V}  \right]\,.
\label{eq:Matter excluded1}
\\[2mm]
\dot{h} +2h^2   &=& \frac{dr_\text{V}}{dy} \,,
\label{eq:Matter excluded2}
\eeqa
\esubeqs
where the last equation corresponds to \eqref{eq:MaxwellFRWdim2}
for $\sigma=1$.
The dimensionless vacuum energy density \eqref{eq:Dimensionless-eV}
for the dimensionless equilibrium chemical potential $u=u_0=-1/3$ is given by
\begin{equation}
r_\text{V}= \frac{1}{2}\,y^2 +\frac{2}{3}\,y^3 +\frac{1}{6}\,y^4 \,,
\label{eq:DimensionlessVacEn2}
\end{equation}
which obviously vanishes in the equilibrium state $y=0$.

In order to simplify the analysis, we, first, consider matter with a
nonzero time-independent EOS parameter,
\beq\label{eq:wM-assumption}
w_\text{M} > 0\,,
\eeq
so that the matter energy density from \eqref{eq:ConservationFRWdim2}
can be neglected asymptotically, as will become clear later on.

\subsection{Vacuum oscillations}
\label{sec:Vacuum-oscillations}

Close to equilibrium, equations  \eqref{eq:Matter excluded1} and
\eqref{eq:Matter excluded2} can be linearized:
\begin{equation}
\ddot{y} +2\dot{h}=0\,,\quad\dot{h}=y\,.
\label{eq:Oscillator}
\end{equation}
The solution of these equations describes rapid oscillations
near the equilibrium point:
\bsubeqs\label{eq:OscillatingSolution}
\beqa
y&=&y_0\,  \sin \omega \tau\,,\qquad\;\;
h=h_0-\frac{y_0}{\omega}\,\cos\omega \tau\,,
\\[2mm]
r_\text{V}&=& \frac{1}{2}\,y_0^2\, \sin^2\omega \tau\,,\quad
\omega^2=2\,.
\eeqa
\esubeqs
The (dimensionless) oscillation period of $y$ and $h$ is given by
\begin{equation}
 \tau_0 = 2\pi/\omega=\pi\sqrt{2}\approx 4.44\,.
\label{eq:OscillatingPeriod}
\end{equation}
The corresponding oscillation period of the vacuum energy density
$r_\text{V}$ is smaller by a factor $2$, so that numerically this period is
given by $\tau_0/2\approx 2.22$.
Both oscillation periods will be manifest in the numerical results of
Sec.~\ref{sec:Numerical results}.

\subsection{Vacuum energy decay}
\label{sec:Vacuum-energy-decay}

The neglected quadratic terms in equations \eqref{eq:Matter excluded1} and
 \eqref{eq:Matter excluded2}  provide
the slow  decay of the amplitudes in \eqref{eq:OscillatingSolution},
namely, the $f$--field oscillation amplitude $y_0(\tau)$,
the Hubble term $h_0(\tau)$,
and the vacuum energy density averaged over fast oscillations
$\langle r_\text{V}\rangle = y_0^2(\tau)/4$.

The explicit behavior is found by expanding the functions $y(\tau)$ and
$h(\tau)$ in powers of $1/\tau$ and keeping terms up to $1/\tau^2$:
\bsubeqs\label{eq:TauExpansion-y-h-ydot-hdot}
\beqa
y=\frac{b(\tau)}{\tau} + \frac{c(\tau)}{\tau^2}\,&,&\quad
h=\frac{l(\tau)}{\tau} + \frac{m(\tau)}{\tau^2}  \,,
\label{eq:TauExpansion-y-h}
\\[2mm]
\dot{y}=\frac{\dot{b}}{\tau} + \frac{\dot{c}-b}{\tau^2}\,&,&\quad
\dot{h}=\frac{\dot{l}}{\tau} + \frac{\dot{m}-l}{\tau^2} \,,
\label{eq:TauExpansion-ydot-hdot}
\\[2mm]
\ddot{y}=\frac{\ddot{b}}{\tau} + \frac{\ddot{c} -2\dot{b}}{\tau^2}\,&,&\quad
\label{eq:TauExpansion-ydotdot}
\eeqa
\esubeqs
where the equality sign has been used rather freely.
Collecting the $1/\tau$ terms, we get homogeneous
linear equations for $b(\tau)$ and $l(\tau)$,
which are actually the same as the linear ODEs \eqref{eq:Oscillator}
with $y$ replaced by $b$ and $h$ replaced by $l$.
The  solution of these equations is given by \eqref{eq:OscillatingSolution}
with the same replacements:
\begin{equation}
b(\tau)=b_0 \,  \sin \omega \tau\,,\quad
l(\tau)=l_0-\frac{b_0}{\omega} \,\cos\omega \tau\,,\quad\omega^2=2\,,
\label{eq:OscillatingSolution2}
\end{equation}
where $l_0$ and $b_0$ are numerical coefficients
which ultimately determine the decay of $h(\tau)$ and $r_\text{V}(\tau)$.

In order to obtain these
coefficients, we must collect the $1/\tau^2$  terms. This leads to
inhomogeneous linear equations for the functions $m(\tau)$ and $c(\tau)$.
The consistency of these equations determines the coefficients  $l_0$ and $b_0$.
It suffices to
keep only the zeroth and first harmonics in the functions $m(t)$ and $c(t)$:
 \begin{equation}
m(\tau) = m^{(1)}\sin\omega \tau \,,\quad c(t) =c^{(0)} + c^{(1)}\cos\omega\tau \,.
\label{eq:harmonics}
\end{equation}
As a result, we obtain the following equations for $m(\tau)$ and $c(\tau)$:
\bsubeqs\label{eq:tausquare1and2}
\beqa
\dot{m} - c &=& \left[l_0 -2l_0^2
                      +\frac{1}{2} \,b_0^2\right]
                  +\frac{b_0}{\omega} \,(4l_0-1) \,\cos \omega \tau \,,
\label{eq:tausquare1}
\\[2mm]
2\dot{m} +\ddot{c}&=&  \left[2l_0 - \frac{3}{2} \,b_0^2
                  -3\,\big(1+w_\text{M}\big)\,\left(l_0^2-\frac{1}{2}\,b_0^2\right)\right]
                  +b_0\,\omega \,(l_0+1) \,\cos\omega \tau  \,.
\label{eq:tausquare2}
\eeqa
\esubeqs
 From the consistency of these equations for the
first harmonics of $m$ and $c$, we obtain
  \begin{equation}
4l_0-1=l_0+1\,,
\label{eq:l0}
\end{equation}
which gives  $l_0=2/3$. Similarly, we find from the zeroth harmonic
of \eqref{eq:tausquare2}
 \begin{equation}
w_\text{M}\,\left(\frac{4}{3} -\frac{3}{2} \,b_0^2\right)\,
=0\,,
\label{eq:b0}
\end{equation}
which, for $w_\text{M}\ne 0$, gives $b_0=2\sqrt{2}/3=\omega\, l_0$.

The above results for the coefficients $l_0$ and $b_0$ hold
for the generic case $w_\text{M}>0$, as stated in \eqref{eq:wM-assumption}.
For the special case $w_\text{M}=0$, inspection of \eqref{eq:3eqsFRWdim2}
shows that the same \emph{Ans\"{a}tze} for $y(\tau)$ and $h(\tau)$
can be used, but with the following coefficients:
\bsubeqs\label{eq:l0b0fM}
\beqa
l_0       &=&2/3\,,\quad b_0=d_\text{M}\;\omega\; l_0\,,
\label{eq:l0b0}\\[2mm]
d_\text{M}&=& 1 +\delta_{w_\text{M},0}\,
              \left(\sqrt{1-(9/4)\,r_{\text{M}\infty}}-1\right)\,,
\label{eq:dM}
\eeqa
\esubeqs
where a Kronecker delta has been employed in the expression for the
damping factor $d_\text{M}$
and the coefficient $r_{\text{M}\infty}$ of the $w_\text{M}=0$
asymptotic energy density $r_\text{M}\sim  r_{\text{M}\infty}/\tau^2$
has been assumed to be less than $4/9$.

Altogether, we have the following behavior of $y(\tau)$, $h(\tau)$,
and $r_\text{V}(\tau)$ for $\tau \to \infty$:
\bsubeqs\label{eq:y+h+eV-Asymptotes}
\beqa
 y &\sim& \frac{2}{3}\;d_\text{M}\;\frac{\sqrt{2}}{\tau}\;\sin \omega\tau \,,
 \label{eq:y-Asymptote}\\[2mm]
 h &\sim& \frac{2}{3}\;\frac{1}{\tau} \;
          \Big( 1-d_\text{M}\; \cos \omega\tau \Big)\,,
 \label{eq:h-Asymptote}\\[2mm]
 r_\text{V} &\sim&
 \frac{4}{9}\;d_\text{M}^2\;\frac{1}{\tau^2}\; \sin^2 \omega\tau\,,
\label{eq:eV-Asymptote}
\eeqa
\esubeqs
with dimensionless frequency $\omega=\sqrt{2}$ and
damping factor $d_\text{M}$ given by \eqref{eq:dM}.
This asymptotic solution has some remarkable properties
(in a different context, the same oscillatory behavior of $h$
has been found in Ref.~\cite{Starobinsky1980}; see also the discussion
in the last paragraph of Sec.~\ref{sec:Gravity-with-F-field}). First,
the solution depends rather weakly on the parameter $w_\text{M}$ of the
matter EOS, which is confirmed by the numerical results of the next subsection.
Second, the average value of the vacuum energy density
decays as $\langle r_\text{V}\rangle \propto 1/\tau^{2}$
and the average value of the Hubble parameter
as $\langle h\rangle  \propto 1/\tau$,
while the average scale parameter increases as
$\langle a(\tau)\rangle   \propto \tau^{2/3}$.
Combined, the average vacuum energy density is found to
behave as $\langle r_\text{V}\rangle \propto 1/a^{3}$,
which is the same behavior as that of CDM in a standard FRW universe,
as will be discussed further in Sec.~\ref{sec:Effective-CDM-like-behavior}.

\subsection{Numerical results}
\label{sec:Numerical results}

\begin{figure*}[p] 
\vspace*{-0.5cm}
\includegraphics[width=.7\textwidth]{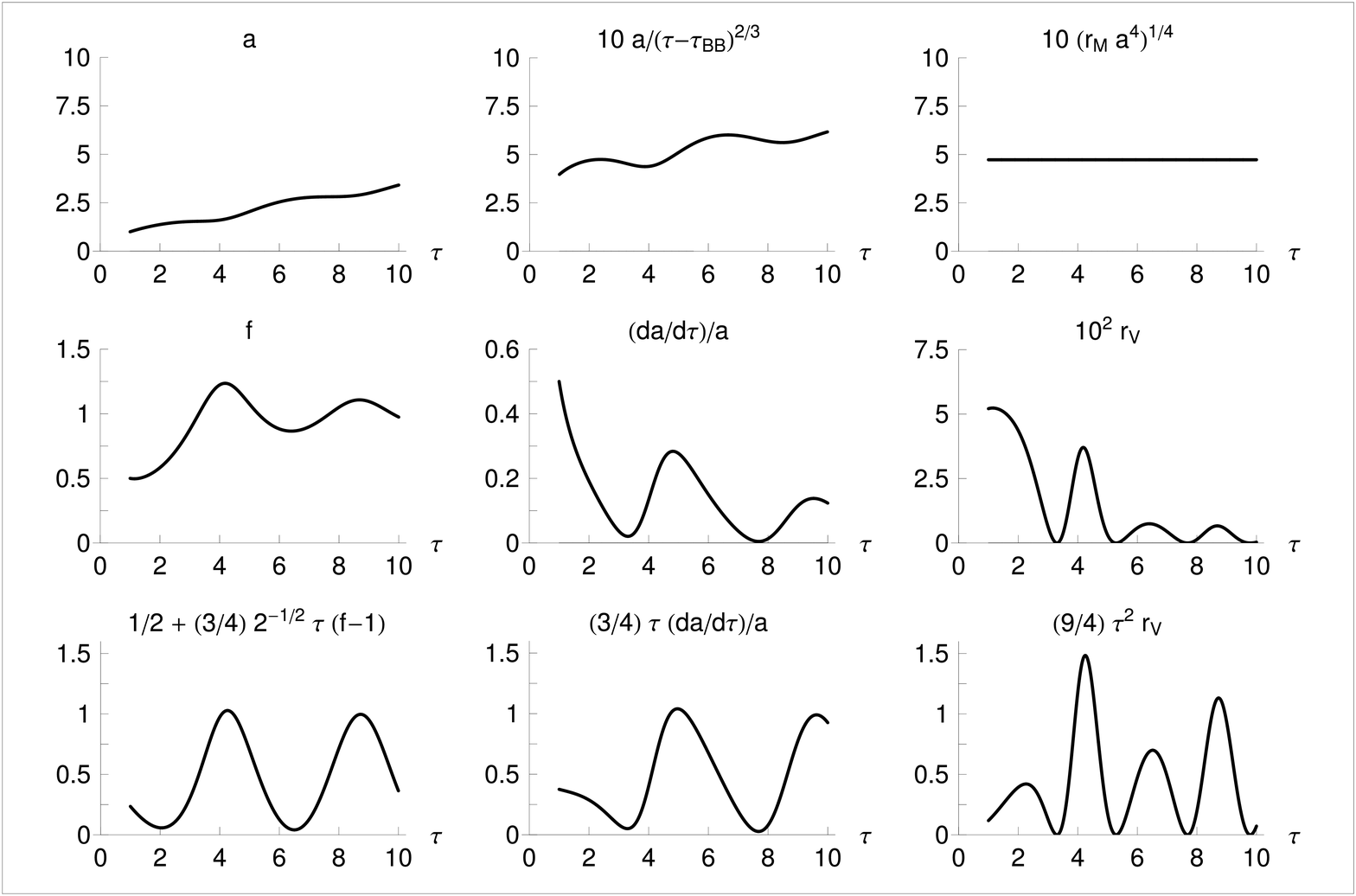}
\caption{Flat FRW universe with scale factor $a(\tau)$,
Hubble parameter $h(\tau) \equiv (da/d\tau)/a$,
ultrarelativistic-matter energy density $\rho_\text{M}(\tau)$,
dynamic vacuum energy density $\rho_\text{V}(F)$
controlled by the vacuum variable $F=F(\tau)$,
and variable gravitational coupling parameter $G=G(F)$.
All variables are scaled to become dimensionless and are denoted by lower-case
Latin letters, for example, $r_\text{V}=r_\text{V}(f)$ and $g=g(f)$.
The specific choices for $r_\text{V}(f)$ and $g(f)$
are given by \eqref{eq:Dimensionless-eV} at chemical potential $u=u_0=-1/3$
and by \eqref{eq:Ginverse-Ansatz}, respectively.
The parameters of the coupled ODEs \eqref{eq:3eqsFRWdim2} are chosen as
$(\sigma,w_\text{M})=(1,\, 1/3)$ and the boundary conditions at $\tau=1$ are
$(a,h,f,r_\text{M})=(1,\, 1/2,\, 1/2,\, 1/20)$.
The effective parameter $\tau_\text{BB}$ in the middle panel of the
top row has been set to the value $-3$.}
\label{fig:FlatFRWuniverse-relmat-short}
\end{figure*}

\begin{figure*}[b]
\vspace*{-0.25cm}
\includegraphics[width=.7\textwidth]{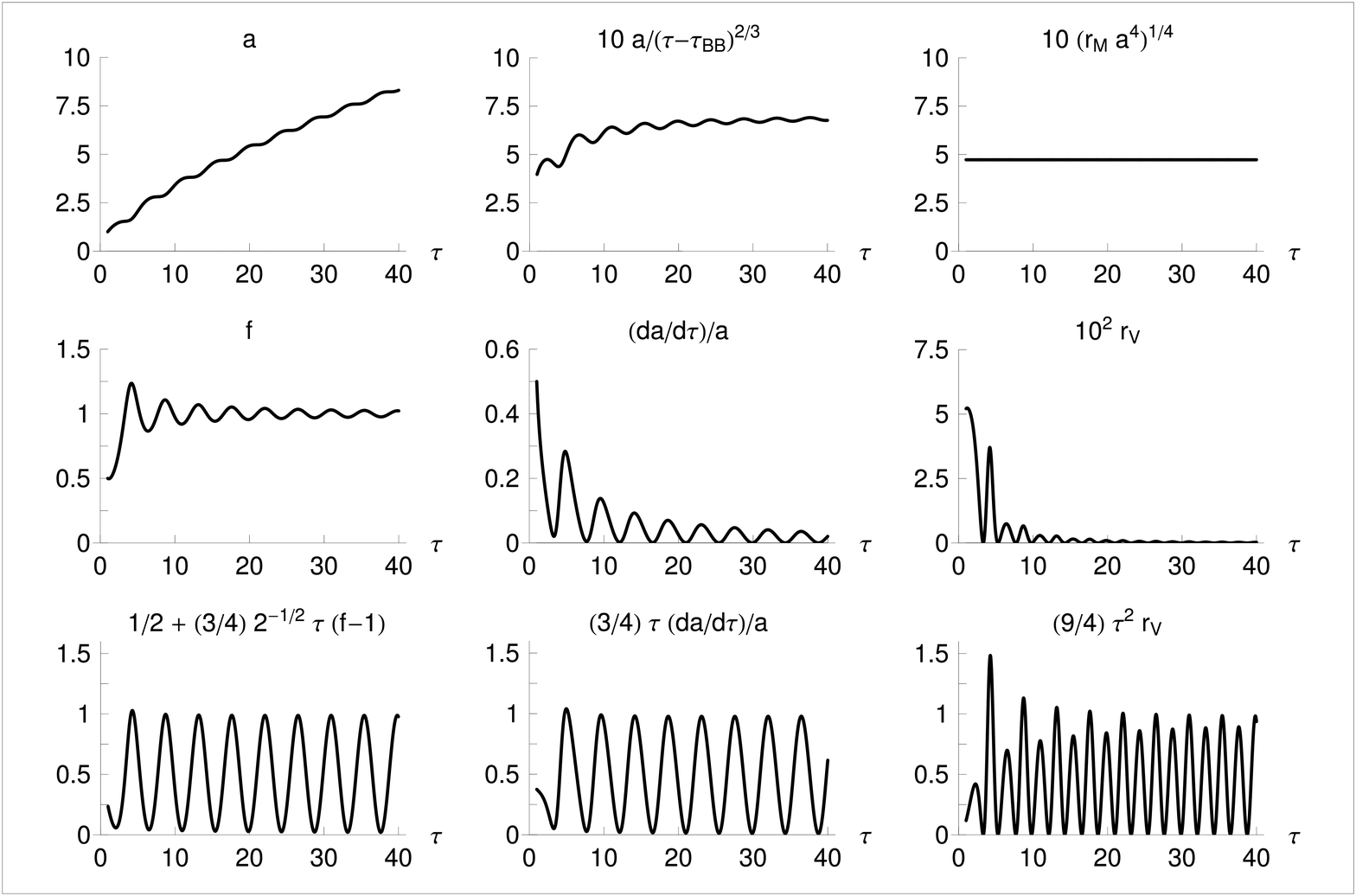}
\caption{Same as Fig.~\ref{fig:FlatFRWuniverse-relmat-short} but
over a longer time.} \label{fig:FlatFRWuniverse-relmat-long}
\end{figure*}

\begin{figure*}[p]
\vspace*{-0.25cm}
\includegraphics[width=.7\textwidth]{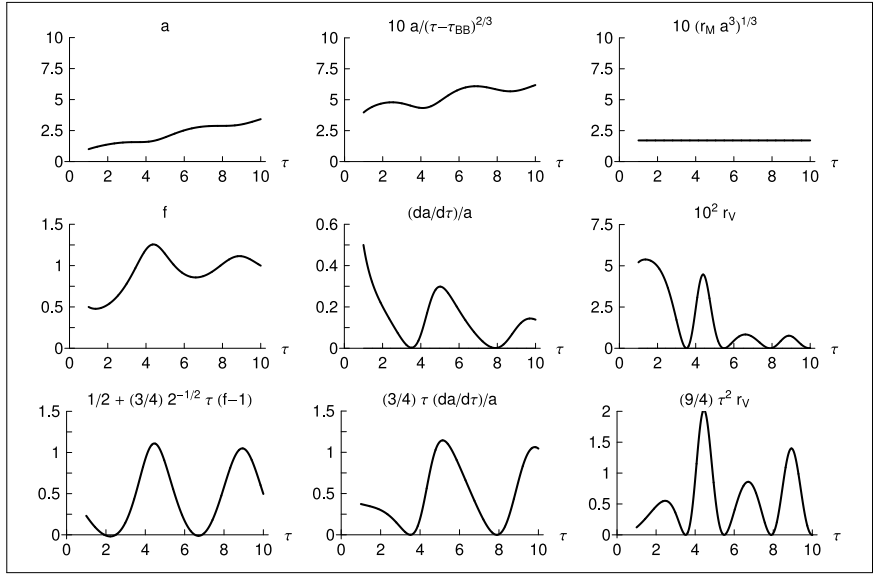}
\caption{Flat FRW universe
with nonrelativistic-matter energy density $r_\text{M}$,
dynamic vacuum energy density $r_\text{V}(f)$ controlled by
the dimensionless vacuum variable $f$,
and variable gravitational coupling parameter $g=g(f)$.
The parameters are chosen as $(u,  \sigma, w_\text{M})=(-1/3,\, 1,\,  0)$
and the boundary conditions at $\tau=1$ are
$(a,h,f,r_\text{M})=(1,\, 1/2,\, 1/2,\, 1/200)$.
The effective parameter $\tau_\text{BB}$ in the middle top-row panel
has been set to the value $-3$.
The matter effects on the oscillation amplitudes of the bottom-row panels
are small, as the damping factor $d_\text{M}$ is close to $1$, namely,
$d_\text{M}\approx 0.986$ from \eqref{eq:dM} with $r_{\text{M}\infty}\approx 0.0125$
[compare to the values of Fig.~\ref{fig:FlatFRWuniverse-nonrelmatHIGH-long}].}
\label{fig:FlatFRWuniverse-nonrelmat-short}
\vspace*{0cm}
\end{figure*}

\begin{figure*}[b]
\vspace*{-0cm}
\includegraphics[width=.7\textwidth]{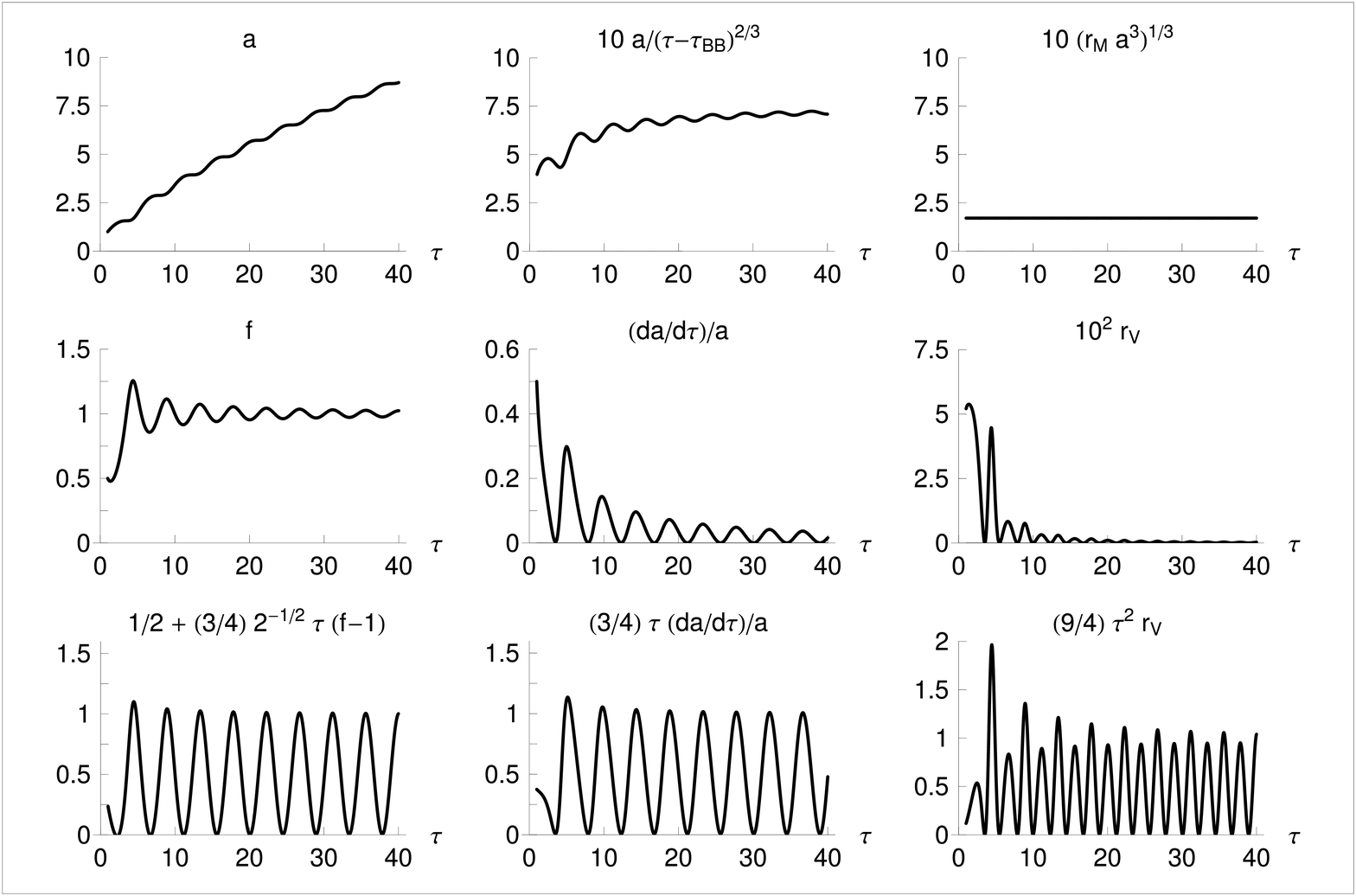}
\caption{Same as Fig.~\ref{fig:FlatFRWuniverse-nonrelmat-short} but
over a longer time.}
\label{fig:FlatFRWuniverse-nonrelmat-long} \vspace*{0cm}
\end{figure*}

\begin{figure*}[p]
\vspace*{-0.25cm}
\includegraphics[width=.7\textwidth]{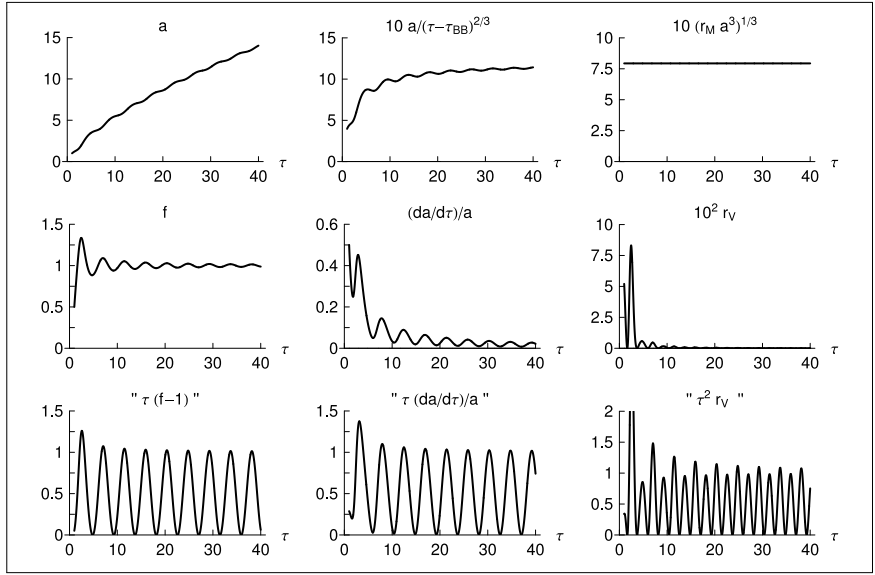}
\caption{Same as Figs.~\ref{fig:FlatFRWuniverse-nonrelmat-short}
and \ref{fig:FlatFRWuniverse-nonrelmat-long}
but with a larger initial density of nonrelativistic matter.
Specifically, the parameters are
$(u,  \sigma, w_\text{M})=(-1/3,\, 1,\,  0)$
and the boundary conditions at $\tau=1$ are
$(a,h,f,r_\text{M})=(1,\, 1/2,\, 1/2,\, 1/2)$.
Plotted on the bottom row are
in the left panel:
$1/2 + (1/d_\text{M})\,( 3/4)\, (1/\sqrt{2}) \,\tau\, (f-1)$,
in the middle panel:
$[(3/2)\, \tau\, (da/d\tau)/a + d_\text{M} - 1]/(2\, d_\text{M})$,
and in the right panel:
$(1/d_\text{M}^2) \, (9/4) \,\tau^2\,r_\text{V}$, for damping factor
$d_\text{M}\approx 0.590$            
from \eqref{eq:dM} with
$r_{\text{M}\infty}\approx 0.290$.   
}
\label{fig:FlatFRWuniverse-nonrelmatHIGH-long}
\vspace*{0cm}
\end{figure*}

\begin{figure*}[b]
\vspace*{-0.25cm}
\includegraphics[width=.7\textwidth]{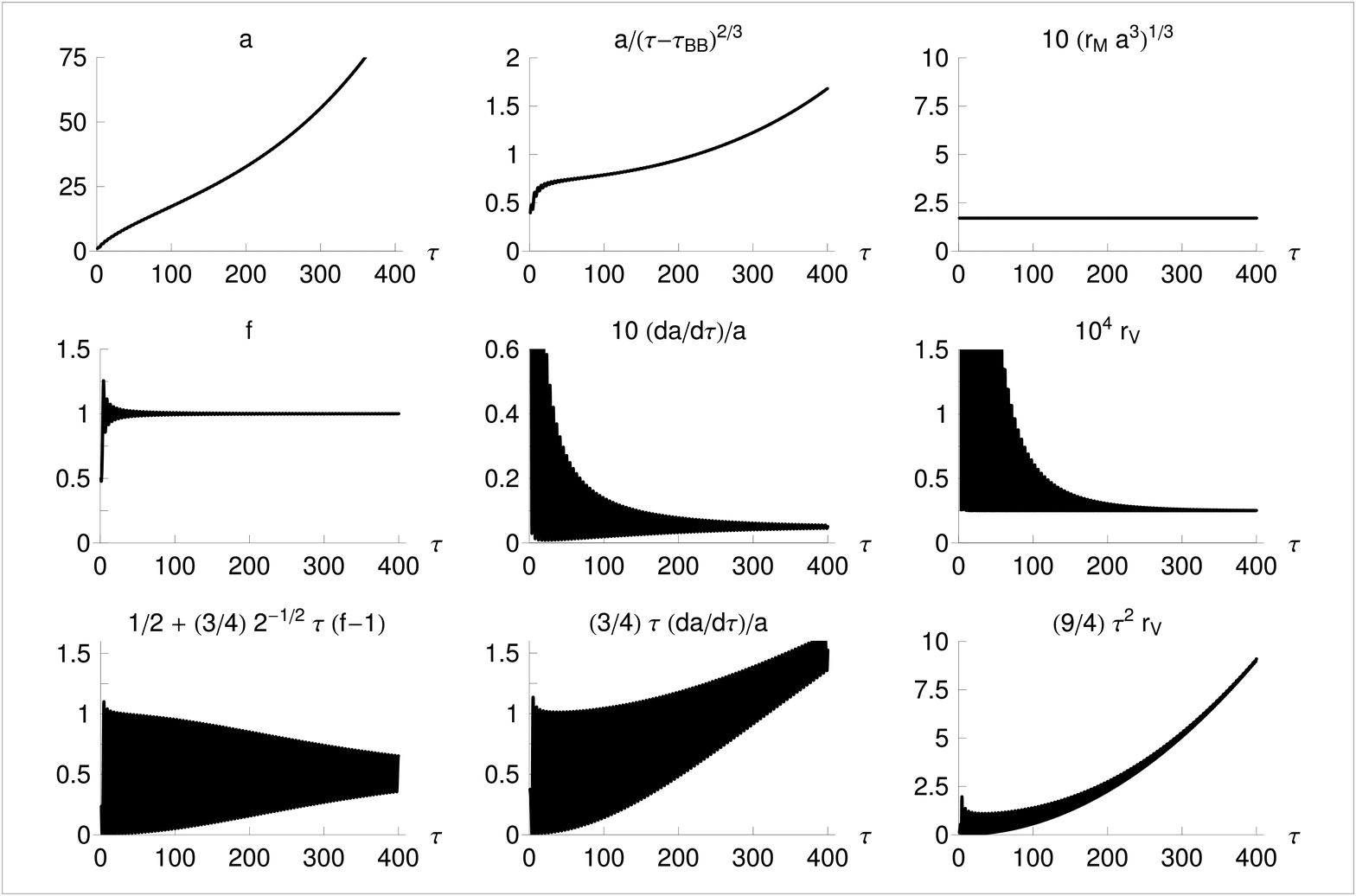}
\caption{Same as Figs.~\ref{fig:FlatFRWuniverse-nonrelmat-short}
and \ref{fig:FlatFRWuniverse-nonrelmat-long}
but with a small perturbation of the dimensionless chemical potential
away from the equilibrium value, $u=u_0+\delta u = -1/3-1/40000$,
and evolved over an even longer time.}
\label{fig:FlatFRWuniverse-deltamu-nonrelmat-long}
\vspace*{0cm}
\end{figure*}

For ultrarelativistic matter ($w_\text{M}=1/3$),
chemical potential $\mu=\mu_0$, and parameter $\sigma=1$,
the numerical solution of the coupled ODEs
\eqref{eq:MaxwellFRWdim2}--\eqref{eq:ConservationFRWdim2}
is given in Figs.~\ref{fig:FlatFRWuniverse-relmat-short}
and \ref{fig:FlatFRWuniverse-relmat-long}.
The behavior near $\tau\sim 1$ is only indicative, as significant
quantum corrections to the classical Einstein equation
can be expected (cf. Sec.~\ref{sec:Equations_mu0}).
Still, the numerical results show clearly that
\begin{enumerate}
\item[(i)]
the equilibrium vacuum is approached asymptotically ($f\to 1$
for $\tau\to \infty$);
\item[(ii)]
the FRW universe (averaged over time intervals larger than
the Planck-scale oscillation period) does not have the
expected behavior $a \propto \tau^{1/2}$ for ultrarelativistic matter but rather
$a \propto \tau^{2/3}\,$;
\item[(iii)]
the same $a \propto \tau^{2/3}$ behavior occurs
if there is initially nonrelativistic matter, as demonstrated by
Figs.~\ref{fig:FlatFRWuniverse-nonrelmat-short} and
\ref{fig:FlatFRWuniverse-nonrelmat-long}
for a relatively small initial energy density
and by Fig.~\ref{fig:FlatFRWuniverse-nonrelmatHIGH-long}
for a relatively large initial energy density;
\item[(iv)]
for a chemical potential $\mu$ slightly different
from the equilibrium value $\mu_0$, the vacuum decay is displayed
in Fig.~\ref{fig:FlatFRWuniverse-deltamu-nonrelmat-long}.
\end{enumerate}
The first three items of the above list of numerical results confirm the
previous asymptotic analytic results of Sec.~\ref{sec:Vacuum-energy-decay}
(these asymptotic results predict, in fact, oscillations between $[0,1]$
for the particular combinations shown on the bottom-row panels of
Figs.~\ref{fig:FlatFRWuniverse-relmat-short}--\ref{fig:FlatFRWuniverse-nonrelmatHIGH-long}),
while the last item shows that, after an initial oscillating stage,
the model universe approaches a de-Sitter stage
(see, in particular, the middle panel of the second row of
Fig.~\ref{fig:FlatFRWuniverse-deltamu-nonrelmat-long}).

\subsection{Effective CDM-like behavior}
\label{sec:Effective-CDM-like-behavior}

The main result of the previous two subsections can be summarized
as follows: the oscillating vacuum energy density $\rho_\text{V}[F(t)]$
and the corresponding oscillating gravitational coupling parameter
$G[F(t)]$ conspire to give the same Hubble expansion as pressureless matter
(e.g., CDM) in a standard FRW universe with fixed gravitational coupling
constant $G=G_\text{N}$. Recall that the standard behavior of the CDM
energy density is given by
$\rho_\text{CDM}(t) \propto a(t)^{-3} \propto t^{-2}$,
which matches the average behavior found in \eqref{eq:y+h+eV-Asymptotes}.

The explanation is as follows. The average values of the rapidly oscillating
vacuum energy density and vacuum pressure act as a source for the slowly
varying gravitational field. The rapidly oscillating parts of $h$ and
$y\equiv f-1$ in the linearized equation \eqref{eq:Oscillator}
correspond to a dynamic system with Lagrangian density
$\half (\dot{y})^2 - \half \omega^2 y^2$ for a time-dependent
homogenous field $y=y(t)$. The $F$ (or $y$) field has no
explicit kinetic term in the action \eqref{eq:actionF}, but derivatives
of $F$ appear in the generalized Einstein Eq. \eqref{eq:EinsteinEquationF2}
via terms with covariant derivatives of $G^{-1}(F)$, which trace back to the
Einstein--Hilbert-like term $R/G(F)$ in \eqref{eq:actionF}.
In a way, the effective Lagrange density
$\half (\dot{y})^2 - \half \omega^2 y^2$ can be said to be induced by gravity.
The pressure of this rapidly oscillating field $y$ is now given
by $P=\half (\dot{y})^2 - \half \omega^2 y^2$.
In turn, this implies that the rapidly oscillating vacuum pressure
is zero on average and that the
main contribution of the oscillating vacuum energy density behaves
effectively as cold dark matter.\footnote{It is known that a
rapidly oscillating homogeneous scalar field in a standard FRW
universe corresponds to pressureless matter
(cf. Sec.~5.4.1 of Ref.~\cite{Mukhanov2005}), but, in our case,
matter plays only a secondary role compared with vacuum energy.
Moreover, the oscillating scalar field gives an oscillating
term in $h(\tau)$ which is subleading (of order $1/\tau^2$), whereas the
oscillating term in \eqref{eq:h-Asymptote} is already of order $1/\tau$.}

Observe that, while the $F$--field itself has an EOS parameter
$w=-1$ corresponding to vacuum energy density,
the net effect of the dampened $F$ oscillations is to mimic
the evolution of cold dark matter with $w=0$ in a standard flat FRW universe.
As mentioned before, this effective EOS parameter $w=0$ is
induced by the interaction of the $F$ and  gravity fields.

An outstanding task is to establish the clustering properties of this type
of oscillating vacuum energy density.
\emph{A priori}, we may expect the same properties as CDM, because the
relevant astronomical length scales are very much larger than the
ultraviolet length scales that determine the microscopic
dynamics of the vacuum energy density. But surprises are, of course,
not excluded.

\subsection{Extrapolation to large times}
\label{sec:Extrapolation}

In Secs.~\ref{sec:Vacuum-energy-decay} and \ref{sec:Numerical results},
we have established that the average vacuum
energy density decreases quadratically with cosmic time.
This behavior follows, analytically, from \eqref{eq:eV-Asymptote}
and, numerically, from the bottom-right panels of
Figs.~\ref{fig:FlatFRWuniverse-relmat-long},
\ref{fig:FlatFRWuniverse-nonrelmat-long},
and \ref{fig:FlatFRWuniverse-nonrelmatHIGH-long}.

Extrapolating this evolution to the present age of the Universe
($t_\text{now}$ $\approx$ $10\;\text{Gyr}$) and
using $|F_0| =s^{-1}\,G^{-1}(F_0) \sim 3/(8\pi G_\text{N})$
for $\sigma \equiv 3s/8\pi=1$,
the numerical value of the average vacuum energy density is given by
\beq
\label{eq:rhoVaverage-now}
\langle \rho_\text{V} (t_\text{now})\rangle \sim \frac{|F_0|}{t_\text{now}^2}
\sim   \frac{E_\text{Planck}^2}{t_\text{now}^2}
=        \left(\frac{t_\text{Planck}}{t_\text{now}}\right)^2\,E_\text{Planck}^4
\approx  \big(4 \times 10^{-3}\;\text{eV}\big)^4\,
         \left(\frac{10^{10}\;\text{yr}}{t_\text{now}}\right)^2,
\eeq
for $t_\text{Planck}=1/E_\text{Planck} \approx 5\times 10^{-44}\;\text{s}$.
The order of magnitude of the above estimate is in agreement with the observed
vacuum energy density of the present universe, which is close to the
critical density of a standard FRW universe
(cf. Refs.~\cite{Weinberg2008,Komatsu-etal2008}
and references therein). If the behavior found had been
$\langle \rho_\text{V}\rangle \propto t^{-n}$ for an integer $n \ne 2$,
this agreement would be lost altogether.
In other words, the dynamic behavior
established in \eqref{eq:eV-Asymptote} is quite nontrivial.

Let us expand on the previous remarks.
For a \emph{standard} flat FRW universe, the \emph{total} energy density is,
of course, always equal to the critical density
$\rho_\text{c} \equiv 3 H^2 /(8\pi G_\text{N})$.
But, here, the gravitational coupling parameter is variable, $G=G(t)$,
and there are rapid oscillations, so that,
for example, $\langle H \rangle^2  \ne \langle H^2 \rangle$.
This explains the following result for the case of
a nonzero matter EOS parameter ($w_\text{M}>0$):
\beq\label{eq:average-Omega-V}
\lim_{t\to\infty}\;
\frac{\langle\rho_\text{V}\rangle}
     {3\langle H \rangle^2 /\big(8\pi\langle G \rangle\big)} = \frac{1}{2}\,,
\eeq
which is of order $1$ but not exactly equal to $1$.
For nonrelativistic matter ($w_\text{M}=0$),
the right-hand side of \eqref{eq:average-Omega-V} is multiplied by a
further reduction factor $d_\text{M}^2= 1-(9/4)\,r_{\text{M}\infty}$,
according to the results of Sec.~\ref{sec:Vacuum-energy-decay}.

Even though the order of magnitude of \eqref{eq:rhoVaverage-now}
or \eqref{eq:average-Omega-V}
appears to be relevant to the observed universe,
the $1/t^2$ behavior of $\langle \rho_\text{V}\rangle $
contradicts the current astronomical data on  ``cosmic
acceleration''~\cite{Frieman-etal2008,Astier-etal2006,Riess-etal2007}.
A related problem is the CDM-like expansion of the model universe,
$a \propto t^{2/3}$, whereas big bang nucleosynthesis requires
radiation-like expansion, $a \propto t^{1/2}$, at least for the
relevant temperature range.
Clearly, there are many other processes that intervene between
the very early (Planckian) phase of the Universe
and later phases such as the nucleosynthesis era and the present epoch.
An example of a relevant process may be particle production
(e.g., by parametric resonance~\cite{Mukhanov2005,Kofman-etal1997}),
which can be expected to be effective because of
the very rapid (but small-amplitude) oscillations.\footnote{Energy
exchange between dark matter and dark energy
(for example, between cold dark matter and dynamic vacuum energy) may,
in fact, be essential to explain
the current epoch of cosmic acceleration; cf. Ref.~\cite{Klinkhamer2008}.
The model considered in the present article
does not allow for energy exchange between dark matter and dark energy,
as \eqref{eq:matter energy-conservation} makes clear.
However, it may be that the effects of such an interaction
can be partially incorporated in our model as a small perturbation
of the chemical potential away from the value \eqref{eq:mu-Ansatz}.
The resulting behavior with exponential expansion setting in at large
times (for small negative perturbations $\delta u$)
is shown in Fig.~\ref{fig:FlatFRWuniverse-deltamu-nonrelmat-long}.}
A further possible source of modified vacuum energy behavior
may be the change of EOS parameter $w_\text{M}= 1/3$ to $w_\text{M}= 0$,
which occurs when the expanding universe leaves the radiation-dominated epoch.
Still, there is a possibility that these and other processes are only secondary
effects and that the main mechanism of dark-energy dynamics at the early
stage is the decay of vacuum energy density by oscillations.

Another aspect of the large-time extrapolation
concerns the variation of Newton's ``constant.''
For the theory \eqref{eq:EinsteinF-all} and
the particular \emph{Ansatz }\eqref{eq:Ginverse-Ansatz},
the gravitational coupling parameter $G(t)$ is found to
relax to an equilibrium value in the following way:
\beq\label{eq:time-dependence-Ginverse}
G^{-1}(t)\sim G^{-1}_\infty \,
              \left(1 + c_0 \,\frac{t_\text{UV}}{t}\,
                    \sin \left(\frac{t}{t_\text{UV}}\right)\right)\,,
\eeq
with $c_0$ a constant of order unity,
$G_\infty$ a gravitational constant presumably very close to the
Cavendish-type value for Newton's constant $G_\text{N}$,
and $t_\text{UV}=\sqrt{\chi  |F_0|/2}$  an ultraviolet timescale
of the order of the corresponding Planckian time scale
$\sqrt{G_\infty \hbar/c^5}$.
The behavior \eqref{eq:time-dependence-Ginverse},
shown qualitatively by the $f$ panels in
Figs.~\ref{fig:FlatFRWuniverse-relmat-short}--\ref{fig:FlatFRWuniverse-deltamu-nonrelmat-long},
is very different from previous suggestions for the dynamics
of $G(t)$, including Dirac's original suggestion $G \propto 1/t$
(cf. Sec.~16.4 of Ref.~\cite{Weinberg1972}).
For the present universe and the solar system in it,
the gravitational coupling parameter \eqref{eq:time-dependence-Ginverse}
would have minuscule oscillations. Combined with the Planck-scale mass of
the $F$ degree of freedom
(cf. the discussion in Sec.~\ref{sec:Effective-CDM-like-behavior}),
this would suggest that all solar-system experimental bounds are satisfied,
but, again, surprises are not excluded.\footnote{After a first version
of the present article was completed, we have become aware of earlier
work on a rapidly varying gravitational coupling parameter $G(t)$;
see, e.g., Refs.~\cite{AccettaSteinhardt1990,SteinhardtWill1994,Perivolaropoulos2003}
and references therein. These articles discuss, in particular,
solar-system experimental bounds and the possibility that the
effective density ratio in a flat FRW universe may be less than
unity, which corresponds to our result \eqref{eq:average-Omega-V} for the case
of dynamic vacuum energy and ultrarelativistic matter with
$\langle\rho_\text{V}\rangle \gg \langle\rho_\text{M}\rangle \geq 0$
asymptotically.}

\vspace*{0mm}
\section{Conclusion}
\label{sec:Conclusion}
\vspace*{0mm}

The considerations of the present article
and its predecessor~\cite{KlinkhamerVolovik2008}
by no means solve the cosmological constant problems,
but may provide hints. Specifically, the new results are
\begin{itemize}
\vspace*{0mm}\item[(i)]
a  mechanism of vacuum-energy decay, which,
starting from a ``natural'' Planck-scale value at very early times,
leads to the correct order of magnitude \eqref{eq:rhoVaverage-now} for
the present cosmological constant;
\vspace*{0mm}\item[(ii)]
the realization from result \eqref{eq:y+h+eV-Asymptotes}
that a substantial part of the inferred CDM
may come from an oscillating vacuum energy density;
\vspace*{0mm}\item[(iii)]
the important role of oscillations of the vacuum variable $q$ (here, $F$),
which drive the vacuum energy density oscillations
responsible for the first two results.
\vspace*{-0mm}
\end{itemize}
Expanding on the last point, another consequence of $q$ oscillations is that
they naturally lead to the creation of hot (ultrarelativistic) matter from the
vacuum. This effective mechanism of energy exchange between vacuum and matter
deserves further study.

\vspace*{0mm}
\section*{\hspace*{-4.5mm}ACKNOWLEDGMENTS}
\vspace*{0mm}
\noindent It is a pleasure to thank A.A. Starobinsky for informative discussions.
GEV is supported in part by the Russian Foundation for
Basic Research (Grant No. 06--02--16002--a) and the Khalatnikov--Starobinsky
leading scientific school (Grant No. 4899.2008.2).

\vspace*{0mm}
\section*{\hspace*{-4.5mm}NOTE ADDED IN PROOF}
\vspace*{0mm}
\noindent Following up on the remarks in the last paragraph of
Sec.~\ref{sec:Gravity-with-F-field},
we have recently shown~\cite{KlinkhamerVolovik2008jetpl}
that, close to equilibrium, the $q$--theory of the quantum vacuum
gives rise to an effective $f(R)$--model
which belongs to the  $R+R^2/M^2$ class of models
with a Planck-scale mass $M\sim E_\text{UV}$.
We have also extended our analysis to a quantum vacuum
containing several conserved $q$--fields, which allows
for the coexistence of different vacua.


\end{document}